\pdfoutput=1
\documentclass[11pt]{article}

\usepackage{acl}
\usepackage{times}
\usepackage{latexsym}
\usepackage[T1]{fontenc}
\usepackage[utf8]{inputenc}
\usepackage{microtype}
\usepackage{inconsolata}
\usepackage{graphicx}
\usepackage{booktabs}
\usepackage{amsmath}
\usepackage{amssymb}
\usepackage{url}

\usepackage{natbib}

\title{Does Financial Speech Need Continual Learning Regularization? A Study of EWC and mini-KFAC on Earnings Call ASR}

\author{Thai Quoc On \\
  Department of Computer Science \\
  Columbia University, NYC, NY, USA \\
  \texttt{tqo2101@columbia.edu} \\
\And
Homayoon Beigi\\
Mechanical Engineering Dept. and\\
Electrical Engineering Dept.\\
Columbia University, NYC, NY, USA\\
and Recognition Technologies, Inc.\\
South Salem, NY, USA\\
 \texttt{homayoon.beigi@columbia.edu}
}

\begin{document}
\maketitle

\makeatletter
\def\@thefnmark{}\@footnotetext{Presented at the 9th International Conference on Natural Language and Speech Processing (ICNLSP 2026), University of Trento, Trento, Italy, September 25--26, 2026.}
\makeatother

\begin{abstract}
Earnings call transcription presents a realistic continual learning scenario where Automatic Speech Recognition (ASR) systems are updated with new recordings but risk catastrophic forgetting of previously learned financial speech patterns. Adapting Whisper-small from a general-domain speech corpus to the Earnings22 benchmark reveals the opposite: all evaluated methods exhibit improvement on held-out financial speech from the same domain during adaptation rather than forgetting. This task complementarity makes regularization counterproductive. Nevertheless, when regularization is warranted, a layer-wise mini-Kronecker-Factored Approximate Curvature (mini-KFAC) approximation seems to achieve lower word error rates (WERs) and more stable optimization than standard diagonal Elastic Weight Consolidation (EWC). The results suggest that task complementarity should be assessed before applying continual learning constraints, and prefer mini-KFAC EWC over diagonal EWC when needed.
\end{abstract}

\section{Introduction}

Financial speech recognition systems face a continual learning challenge in which models must be updated as new data arrives, raising concerns about catastrophic forgetting during sequential fine-tuning \citep{kirkpatrick2017overcoming, rusu2016progressive}.
However, when tasks share acoustic and linguistic structure, adaptation to one partition may improve performance on another, a phenomenon referred to here as task complementarity. Whether catastrophic forgetting actually occurs in financial speech Automatic Speech Recognition (ASR) remains underexplored. Elastic Weight Consolidation (EWC) \citep{kirkpatrick2017overcoming} constrains parameter updates via Fisher information, but diagonal approximations ignore parameter correlations that may be important in large transformer-based ASR models such as Whisper \citep{radford2022robust}. Although \citet{mcinerney2021kronecker} showed that Kronecker-factored EWC substantially reduces forgetting relative to diagonal EWC, this question has not been examined in financial speech.

This work adapts Whisper-small from a TEDLIUM-derived corpus \citep{rousseau2012tedlium, gandhi2023tedlium} to the Earnings22 benchmark \citep{delrio2022earnings22}, comparing unregularized fine-tuning, diagonal EWC, and a memory-efficient Kronecker-factored variant: mini-Kronecker-Factored Approximate Curvature EWC (mini-KFAC EWC). The central finding is that all methods show statistically significant negative forgetting, with performance on held-out financial speech improving during adaptation rather than degrading, within this single-domain setting. These findings contribute two practical guidelines to the continual learning literature: task similarity assessment should precede any regularization decision, and Kronecker-factored approximations are preferable to diagonal Fisher methods when constraints are necessary.

\section{Related Work}

Continual learning studies how to learn sequential tasks without forgetting prior knowledge. Specifically, EWC \citep{kirkpatrick2017overcoming} addresses this by penalizing parameter drift using Fisher information, while replay and architectural expansion methods offer alternative approaches \citep{rebuffi2017icarl, rusu2016progressive, zenke2017continual, serra2018lifelong}.

ASR continual learning remains less studied than vision \citep{vander2022continual, plantinga2023continual}, and financial speech benchmarks such as Earnings22 \citep{delrio2022earnings22} are underexplored in this context.

Kronecker-Factored Approximate Curvature (KFAC) approximates the Fisher information layer-wise, as $F \approx G \otimes A$ where $A$ is input covariance, $G$ is gradient covariance, and $\otimes$ denotes the Kronecker
product~\cite{martens2015optimizing}, capturing parameter dependencies more effectively than diagonal methods while remaining scalable \citep{martens2015optimizing}. Prior work has shown the effectiveness of Fisher-based regularization in other domains. For example, \citet{mcinerney2021kronecker} reported that Kronecker-factored EWC reduced forgetting by 51\%, compared with 27\% for diagonal EWC, on BERT-based medical entity linking. Additionally, \citet{ahadzi2025continual} applied EWC to Whisper-small for children's ASR and achieved a 5.21\% relative Word Error Rate (WER) reduction over unregularized fine-tuning. However, neither work examines Kronecker-factored approximations for ASR nor the financial speech domain, and neither reports the task complementarity regime.

\section{Experimental Setup}

\begin{figure}[h]
\centering
\includegraphics[width=\linewidth]{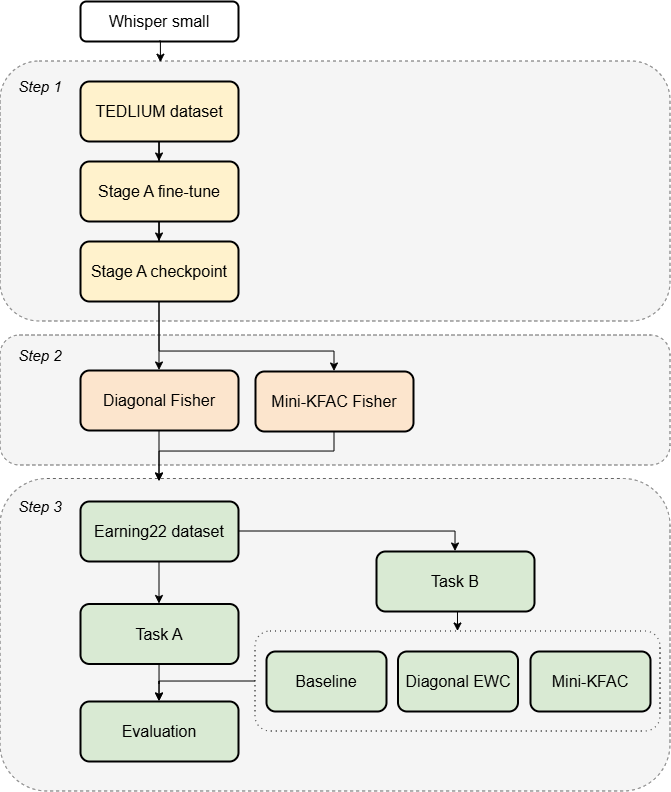}
\caption{The architecture of the continual learning experimental pipeline.}
\label{fig:pipeline}
\end{figure}

\textbf{Model and Data:} Whisper-small \citep{radford2022robust} (244M parameters), a transformer encoder-decoder model pre-trained on large-scale weakly supervised speech data, is used. The experimental pipeline in Figure~\ref{fig:pipeline} proceeds in three steps and mimics a realistic incremental deployment, in which a model is first adapted to a general-domain corpus and then fine-tuned on financial speech, with retention measured on a disjoint partition of the target domain. Particularly, in \textbf{Step~1 (Stage~A)}, the model is adapted on the training split of TEDLIUM-derived speech corpus \citep{rousseau2012tedlium, gandhi2023tedlium}, a large corpus of aligned TED talk speech, to provide a general-domain initialization. In \textbf{Step~2 (Fisher Estimation)}, the Fisher statistics are estimated from Stage~A batches using both diagonal and Kronecker-factored (mini-KFAC) approximations, capturing parameter importance before financial-domain adaptation begins. In \textbf{Step~3 (Stage~B)}, the model is fine-tuned on the Earnings22 data \citep{delrio2022earnings22}, which is partitioned into two disjoint subsets through MD5 hashing of file identifiers: \textbf{Task~B}, the target partition used for Stage~B fine-tuning, and \textbf{Task~A}, a held-out retention partition on which the model is never trained, used to evaluate whether prior performance is preserved during the Stage~B adaptation. Since each file identifier corresponds to a distinct earnings call recording, this partitioning guarantees non-overlapping speaker groups across Task~A and Task~B. All three methods: unregularized baseline, diagonal EWC, and mini-KFAC EWC are evaluated on both Task~A and Task~B partitions. Task construction rationale and hyperparameter details are in Appendix~\ref{app:task_rationale} and~\ref{app:hyperparams}, and the full code implementation is available at \url{https://anonymous.4open.science/r/continual-asr-finance-508B/}.

\textbf{Methods:}
\begin{itemize}
    \item {Baseline:} Unregularized fine-tuning.
    \item {Diagonal EWC:} Parameter-wise $\ell_2$ regularization \citep{kirkpatrick2017overcoming}:
    $\mathcal{L}_{\text{total}} = \mathcal{L}_{\text{ASR}} + \frac{\lambda}{2}\sum_i F_i(\theta_i - \theta_i^*)^2$,
    where $\mathcal{L}_{\text{ASR}}$ is the ASR loss, $\theta_i$ and $\theta_i^*$ are the current and Stage~A parameters, $F_i$ is the diagonal Fisher estimate for parameter $i$, and $\lambda$ is the regularization strength.
    
    \item {Mini-KFAC EWC:} Layer-wise Kronecker-factored Fisher regularization with per-layer adaptive scaling
    $s_\ell = \sqrt{\bar{\mathrm{Tr}}(A_\ell) \cdot \bar{\mathrm{Tr}}(G_\ell)}$,
    where $A_\ell$ and $G_\ell$ are the Kronecker factors (input activation and output gradient covariances) for layer $\ell$, and $\bar{\mathrm{Tr}}(\cdot)$ is the mean trace (trace divided by matrix dimension); $s_\ell$ is clipped to $[1.0,\, 5.0]$.
\end{itemize}

Mini-KFAC was selected due to the computational cost of full KFAC. This approach was termed mini-KFAC to reflect three memory-efficient approximations relative to the original formulation \citep{martens2015optimizing}: restriction to linear layers with dimension $\leq$ 2048, row subsampling of activation and gradient matrices to at most 100 rows per batch, and CPU-side accumulation of the Kronecker factors. Additional rationale details are provided in
Appendix~\ref{app:hyperparams}. 
In this set-up, a unified layer-wise regularization scheme is applied  across all continual learning methods such that encoder layers receive stronger regularization ($\lambda_{\text{enc}} = 10{,}000$)
than decoder layers ($\lambda_{\text{dec}} = 2500$) given the encoder's greater
acoustic specificity \citep{vander2022continual}. The lower global scale for
mini-KFAC against EWC ($\lambda_{\text{scale}} = 48$ vs.\ $10{,}000$) compensates for the
larger magnitude of matrix quadratic-form penalties relative to scalar
per-parameter terms (Appendix~\ref{app:hyperparams}). Moreover, coefficients were selected empirically to keep the EWC regularization influential but non-dominant, with regularization strength ramped linearly from zero over the first 50 steps (Appendix~\ref{app:gradients}).

\textbf{Evaluation Metrics:} Three metrics are reported: (1) The Task~B WER, which is the best word error rate on the Task~B partition during fine-tuning; (2) The Task~A WER, signifying the word error rate on the Task~A retention partition after fine-tuning completes; and (3) The forgetting rate, defined as the change in the Task~A WER before and after the Stage~B fine-tuning, i.e.\ $\text{Forgetting} = \text{WER}_A^{\text{final}} - \text{WER}_A^{\text{pre-B}}$, where pre-B is the Task~A WER evaluated immediately after the Stage~A training and before any Stage~B updates. Negative forgetting indicates that Task~A performance improved during Stage~B training rather than degraded. All results are reported as mean $\pm$ standard deviation across five seeds (3, 17, 24, 35, and 50).

\section{Results}

\begin{table}[t]
\centering
\caption{Results across five seeds (mean $\pm$ std). Task~B WER is the best validation WER on the fine-tuning partition. Task~A WER is evaluated on the best Task~B checkpoint. Forgetting is the change in Task~A WER relative to pre-B; negative values indicate improvement. Superscripts indicate statistical significance. $^\dagger$ denotes a one-sample, one-tailed $t$-test showing that forgetting is significantly below zero ($p < 0.05$), see Appendix~\ref{app:stats}.}
\label{tab:results}
\small
\resizebox{\columnwidth}{!}{
\begin{tabular}{lccc}
\toprule
\textbf{Method} & \textbf{Task~B WER} & \textbf{Task~A WER} & \textbf{Forgetting} \\
 & \textbf{(\%)} & \textbf{(\%)} & \textbf{(\%)} \\
\midrule
Baseline       & $31.94 \pm 1.73$ & $20.68 \pm 1.63$ & $-4.87 \pm 2.47^\dagger$ \\
EWC (Diagonal) & $35.23 \pm 4.91$ & $21.88 \pm 2.08$ & $-3.66 \pm 1.95^\dagger$ \\
Mini-KFAC EWC  & $34.26 \pm 3.57$ & $21.52 \pm 2.17$ & $-4.02 \pm 2.41^\dagger$ \\
\bottomrule
\end{tabular}
}
\end{table}

\subsection{Finding 1: Task Complementarity Makes Regularization Counterproductive}

All three methods exhibit statistically significant negative forgetting: Task~A WER improves during Stage~B fine-tuning for every method (one-sample $t$-tests: baseline $t=-4.41$, $p=0.006$; EWC $t=-4.20$, $p=0.007$; mini-KFAC $t=-3.73$, $p=0.010$; all one-tailed, $df=4$). The stronger the regularization constraint, the less improvement is permitted: baseline forgetting is $-4.87\%$, mini-KFAC $-4.02\%$, and diagonal EWC $-3.66\%$. This pattern indicates that earnings call speech shares acoustic and linguistic patterns across both partitions, so regularization prevents beneficial co-adaptation rather than blocking genuine interference, making it counterproductive in this setting.



\subsection{Finding 2: Unregularized Fine-tuning Remains Superior}

\begin{figure}[t]
\centering
\includegraphics[width=\linewidth]{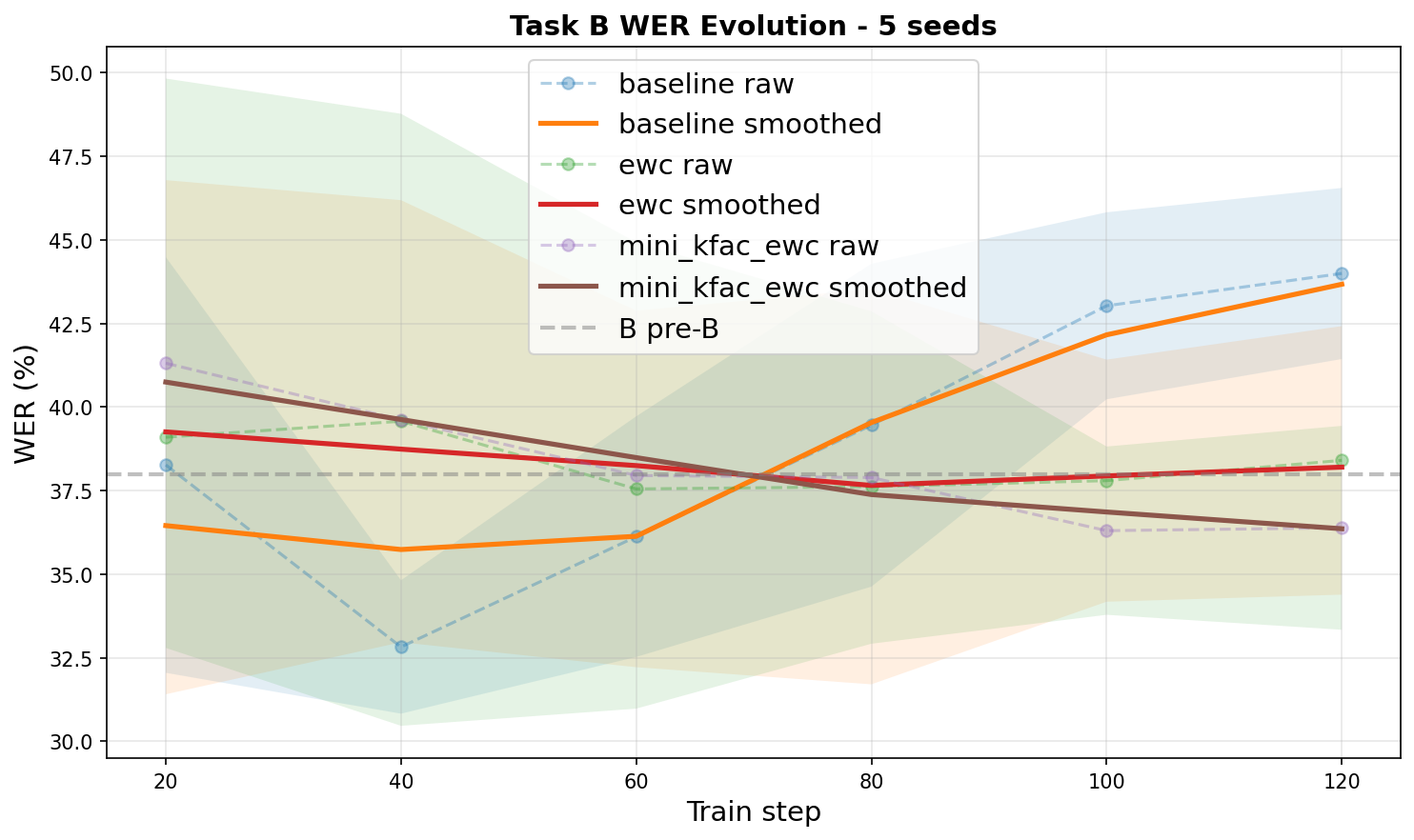}
\caption{Task~B WER evolution during Stage~B fine-tuning.}
\label{fig:task_b_wer}
\end{figure}

Despite the advantage of mini-KFAC over diagonal EWC, baseline fine-tuning achieves lower WER than both regularized methods on Task~B ($31.94\%$ vs.\ $35.23\%$ for EWC and $34.26\%$ for mini-KFAC). Although pairwise $t$-tests are not statistically significant at $n=5$, the effect sizes for Task~B WER are large (baseline vs.\ EWC: $d=0.89$; baseline vs.\ mini-KFAC: $d=0.83$), indicating a practically meaningful pattern that more seeds would likely confirm. This baseline advantage is visible in Figure~\ref{fig:task_b_wer}, where the baseline achieves the lowest WER throughout training.
\begin{figure}[t]
\centering
\includegraphics[width=\linewidth]{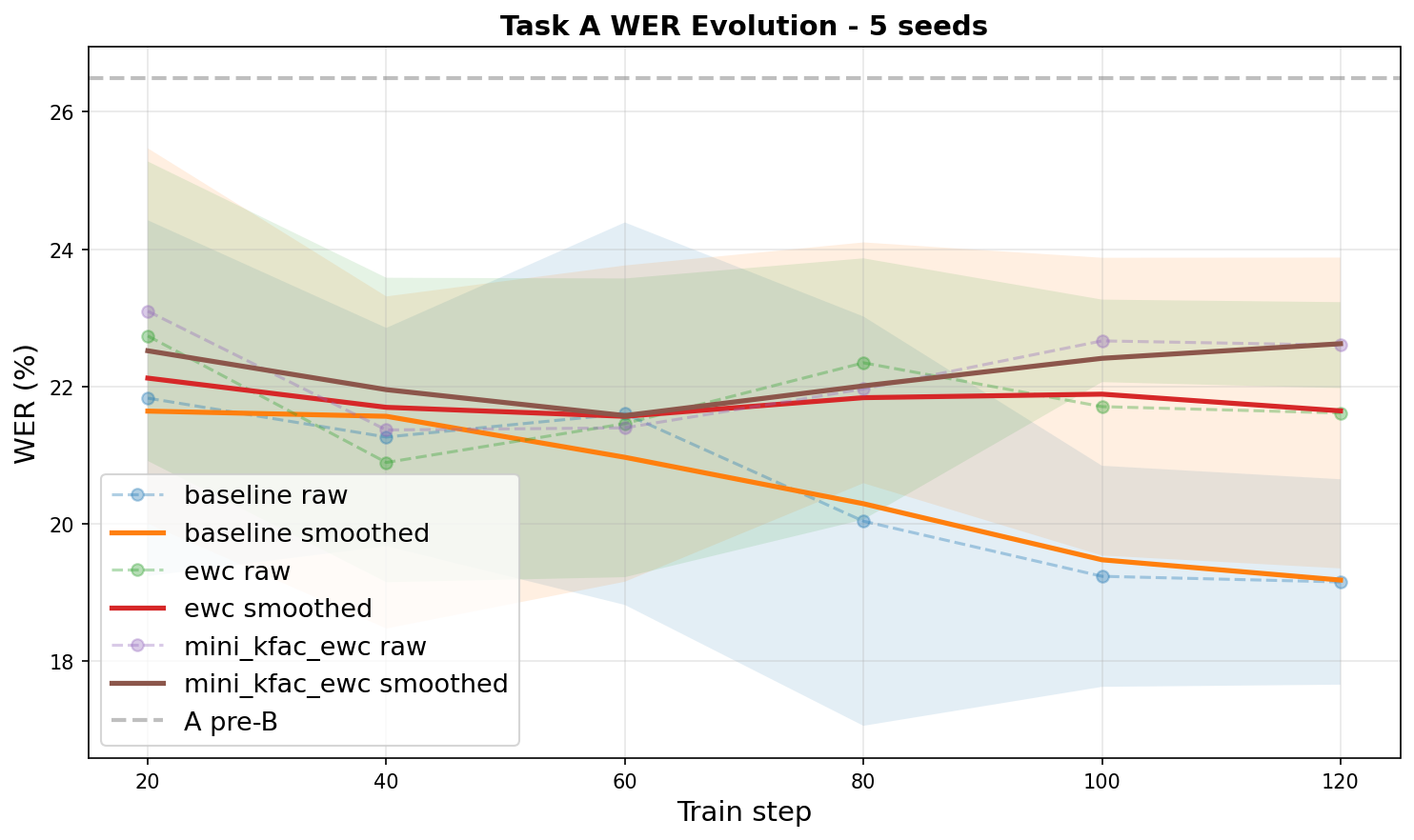}
\caption{Task~A WER evolution during Stage~B fine-tuning.}
\label{fig:task_a_wer}
\end{figure}

\subsection{Finding 3: Mini-KFAC Shows Consistent but Statistically Modest Advantages over Diagonal EWC}

When regularization is applied, mini-KFAC EWC achieves lower WER than diagonal EWC across all metrics (Table~\ref{tab:results}): a $0.97\%$ Task~B WER reduction 
($34.26\%$ vs.\ $35.23\%$) and a $0.36\%$ Task~A WER reduction ($21.52\%$ vs.\ $21.88\%$). Mini-KFAC also shows lower variance across seeds (Task~B std: $3.57$ vs.\ $4.91$), indicating more stable optimization. Figures~\ref{fig:task_a_wer} and~\ref{fig:task_b_wer} show that mini-KFAC converges faster and to a lower WER than diagonal EWC on both tasks across all seeds. However, Welch's two-sample $t$-tests indicate that these differences are not statistically significant at $n=5$ seeds (Task~A: $t=-0.27$, $p=0.80$; Task~B: $t=-0.36$, $p=0.73$; effect sizes $d \approx 0.17$--$0.23$, small; see Appendix~\ref{app:stats}).
The consistent directional advantage of mini-KFAC across all metrics and seeds suggests a genuine but modest benefit of Kronecker factorization: by capturing within-layer parameter correlations, mini-KFAC allows more selective constraints than purely diagonal penalties, though larger-scale experiments are needed to confirm this statistically.

\section{Discussion}

\textbf{Interpreting the absence of forgetting:} The lack of forgetting is not merely a consequence of the task construction. Although Task~A and Task~B come from the same dataset, meaningful interference between disjoint partitions would still be expected to cause degradation. Instead, the consistent improvement on Task~A suggests that positive transfer dominates in this setting. This highlights the importance of assessing task relevance before applying continual learning regularization, as strong transfer can render such constraints unnecessary or even detrimental.

\textbf{Scope of the ``forgetting'' framing:} The phrase ``negative forgetting'' is used in this work as a label for the signed WER change defined in Section~3, not as a claim of resistance to forgetting under genuine domain shift. Task~A and Task~B are both Earnings22 partitions from the same financial-speech distribution, so the Task~A improvement reflects fine-tuning generalizing across speakers within one domain, rather than the classical retention-adaptation trade-off studied under larger domain shifts. This within-domain design was chosen deliberately (Appendix~\ref{app:task_rationale}): it isolates regularization effects from confounding domain shift, and represents the realistic setting of incremental updates from the same data distribution. Because the two tasks are closely related, the setup approximates a best case for avoiding forgetting, and the results show that regularization is unnecessary and counterproductive here. The results say nothing about what happens under larger domain shift, a question discussed in Appendix~\ref{app:when_reg_works}.

\textbf{Sources of task complementarity:}
The two corpora draw from non-overlapping speaker populations: TEDLIUM encompasses 228 named TED conference speakers \citep{gandhi2023tedlium}, distinct from global corporate executives and financial analysts in Earnings22 \citep{delrio2022earnings22}. Both consist of prepared or semi-prepared professional English speech with controlled recording conditions that provide cleaner signal than far-field or telephone-quality speech \citep{gandhi2023tedlium, delrio2022earnings22}. These shared properties are consistent with the encoder representations acquired during Stage~A generalizing constructively to Earnings22, and offer a plausible account of the negative forgetting observed across all methods.

\textbf{Mini-KFAC outperformance over diagonal EWC:} Diagonal Fisher treats parameters independently, producing overly restrictive constraints in high dimensions. Combined with large but effective regularization strength, this makes diagonal EWC rigid. In contrast, Mini-KFAC uses a Kronecker factorization that captures layer-wise structure, allowing more flexible updates that better preserve second-order geometry. Its layer-wise scaling further corrects Fisher magnitude imbalance across the network, which diagonal EWC cannot address.

\textbf{Practical recommendations}: Practitioners should include evaluating forgetting before applying continual learning regularization as stable or negative forgetting indicates task complementarity and reduces the need for additional constraints. When catastrophic forgetting is observed, mini-KFAC EWC is preferable to diagonal EWC. Moreover, careful tuning of regularization strength is also important, as overly restrictive constraints can limit adaptation and degrade performance, while overly loose or small values can provide insufficient penalty to preserve prior knowledge, making the regularization ineffective.

\section{Conclusion}

In this study, a comparative evaluation of Fisher-based continual learning methods is presented for financial speech recognition, using the Whisper-small model. The presented results suggest that preserving layer-wise parameter interactions through Kronecker-factored approximations may provide more effective regularization than diagonal Fisher approximations in this setting. Furthermore, all methods exhibit negative forgetting, with Task~A performance improving throughout the Stage~B fine-tuning process. This observation suggests strong task complementarity in the financial speech domain and, in this within-domain setting, contrasts with the commonly expected retention–adaptation trade-off in continual learning, where adaptation to a new task often degrades performance on previously learned tasks. These findings establish a practical guideline: task similarity should be assessed before applying regularization, and when regularization is warranted, mini-KFAC EWC might be the better choice.

\section{Limitations}

This study uses a controlled two-partition Earnings22 setup and reports results over five random seeds. While five seeds yield statistical significance of negative forgetting ($p < 0.05$ across all methods, with two of three methods at $p < 0.01$), pairwise comparisons between regularization strategies remain underpowered at this scale (effect sizes $d \leq 0.23$ for mini-KFAC vs.\ diagonal EWC). These tests also rest on a normality assumption that is difficult to verify at $n=5$ and Appendix~\ref{app:stats} discusses this limitation and notes bootstrap-based alternatives as a more robust next step. Larger seed counts and longer task sequences remain important directions for future work to confirm the relative advantage of Kronecker-factored regularization. Because Task~A and Task~B are both partitions of the same Earnings22 corpus (Appendix~\ref{app:task_rationale}), the negative forgetting observed here is a within-domain result and should not be read as evidence that regularization is unnecessary under larger domain shifts. Additionally, scenarios where regularization may be beneficial are discussed in Appendix~\ref{app:when_reg_works}.

\bibliography{custom}

\appendix



\section{Task Construction Rationale}
\label{app:task_rationale}

Task~A and Task~B are disjoint Earnings22 partitions with different speakers, and
recordings, enabling controlled evaluation of retention within the same financial-speech domain.
This study uses within-domain partitions rather than cross-domain tasks for two reasons.
First, this isolates the effect of Fisher-based regularization from the confounding influence of
large distributional shifts and any forgetting measured here is directly attributable to parameter drift rather than domain mismatch. Second, this represents a realistic deployment scenario: production earnings-call systems are updated incrementally as new quarterly recordings become available, where successive batches are drawn from the same underlying financial-speech distribution. This controlled design provides a lower-bound study of catastrophic forgetting. Stronger domain shifts (e.g., Earnings22 followed by a medical or conversational corpus) may produce different regularization dynamics, and should be considered as an important direction for future work. Additionally, Earnings22 is used rather than Earnings21 \citep{delrio2021earnings21} to avoid manual segmentation steps that could confound the continual learning analysis.

\section{Hyperparameter Details}
\label{app:hyperparams}
All experiments were conducted on a single NVIDIA L4 GPU. Hyperparameters and settings were selected to fit within the available GPU memory and runtime budget while maintaining stable optimization. Full training configuration:
\begin{itemize}
    \item Learning rate (Stage B): $3 \times 10^{-6}$
    \item Training steps: 120 steps (Stage A: 100, Stage B: 120), where one step corresponds to a single forward and backward pass on one mini-batch
    \item Evaluation interval: every 20 training steps (10 optimizer updates)
    \item Gradient accumulation: 2 steps per optimizer update (effective batch size 8)
    \item Batch size: 4 samples per training step
    \item Early stopping patience: 6 evaluations
    \item Warmup ratio: 10\% of total steps
    \item EWC ramp-up: 50 steps (linearly scale $\lambda$ from 0 to target)
    \item Fisher computation: 800 batches
\end{itemize}
Regularization coefficients:
\begin{itemize}
    \item Diagonal EWC: $\lambda_{\text{base}}=5000$, $\lambda_{\text{enc}}=10{,}000$, $\lambda_{\text{dec}}=2500$; global scale $\lambda_{\text{scale}}=10{,}000$
    \item Mini-KFAC EWC: same base coefficients; global scale $\lambda_{\text{scale}}=48$
\end{itemize}

Dataset partitioning: Earnings22 recordings are deterministically split into disjoint Task~A and Task~B subsets via MD5 hashing of file identifiers. Stage~B fine-tuning uses only Task~B; Task~A is never seen during training and is used solely to measure retention performance.


\textbf{Regularization scale rationale:} The large disparity between the diagonal EWC global scale ($\lambda_{\text{scale}} = 10{,}000$) and the mini-KFAC global scale ($\lambda_{\text{scale}} = 48$) is intentional and not a tuning artifact, where $\lambda_{\text{scale}}$ is the global multiplier applied to the EWC penalty term before it is added to the ASR loss. The diagonal EWC penalty sums scalar
Fisher estimates over individual parameters; each term is small. Specifically, the diagonal EWC penalty takes the form
$\frac{\lambda}{2} \sum_i F_i (\theta_i - \theta_i^*)^2$,
where $F_i \in \mathbb{R}$ is the scalar diagonal Fisher estimate for parameter $i$, $\theta_i \in \mathbb{R}$ is the current value of parameter $i$, and $\theta_i^* \in \mathbb{R}$ is its value at the end of Stage~A; each term is
a scalar and the sum grows only linearly with the number of parameters. The mini-KFAC penalty instead computes a matrix quadratic form $\Delta W_\ell^\top (G_\ell \otimes A_\ell) \Delta W_\ell$ over the full weight matrix of each layer, which is naturally orders of magnitude larger in raw value, where $\Delta W_\ell = W_\ell - W_\ell^* \in \mathbb{R}^{p
\times q}$ is the weight change matrix for layer $\ell$ (current minus Stage~A weights), $G_\ell \in \mathbb{R}^{p \times p}$ is the output gradient covariance, $A_\ell \in \mathbb{R}^{q \times q}$ is the input activation covariance,
$\otimes$ denotes the Kronecker product, and the resulting quadratic form aggregates correlations across all $p \times q$ parameters of the layer simultaneously rather than summing independent scalar terms. Both $\lambda_{\text{scale}}$ coefficients were set so that the gradient ratio
$\|\nabla_\theta \mathcal{L}_{\text{EWC}}\| / \|\nabla_\theta
\mathcal{L}_{\text{ASR}}\|$ remains in a balanced, non-dominating range throughout Stage~B training, where $\nabla_\theta \mathcal{L}_{\text{EWC}}$
is the gradient of the regularization term with respect to all model parameters $\theta$, $\nabla_\theta \mathcal{L}_{\text{ASR}}$ is the corresponding gradient
of the ASR loss, and a ratio well below 1 ensures the constraint guides but does not overwhelm task learning; this was verified empirically (Appendix~\ref{app:gradients}).

\textbf{Mini-KFAC selection rationale:} Computing the full Fisher matrix for a 244M-parameter model is computationally intractable. Mini-KFAC addresses this by approximating it layer-wise as a Kronecker product of input activation and output gradient covariances. This is computed using subsampled activations and gradient statistics extracted during GPU-based forward and backward passes, while Fisher accumulation is performed on the CPU to reduce GPU memory overhead. This hybrid design enables efficient estimation over approximately 800 batches within a single-GPU training budget, while still capturing inter-parameter correlations that purely diagonal methods ignore.

\section{Training Dynamics Analysis}
\label{app:loss_breakdown}
\begin{figure}[t]
\centering
\includegraphics[width=\linewidth]{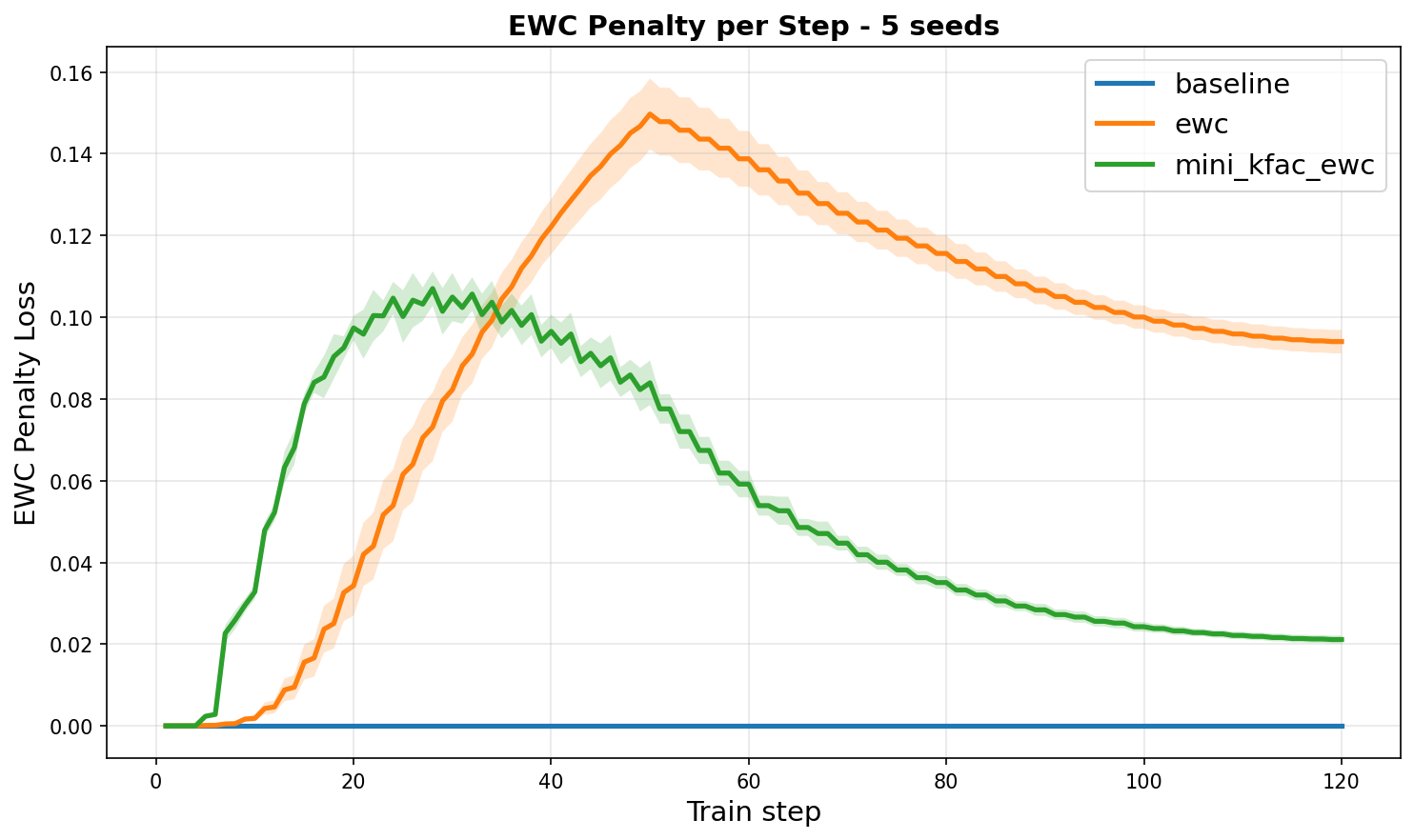}
\caption{EWC penalty evolution during Stage~B adaptation.}
\label{fig:penalty}
\end{figure}

\begin{figure}[h]
\centering
\includegraphics[width=\linewidth]{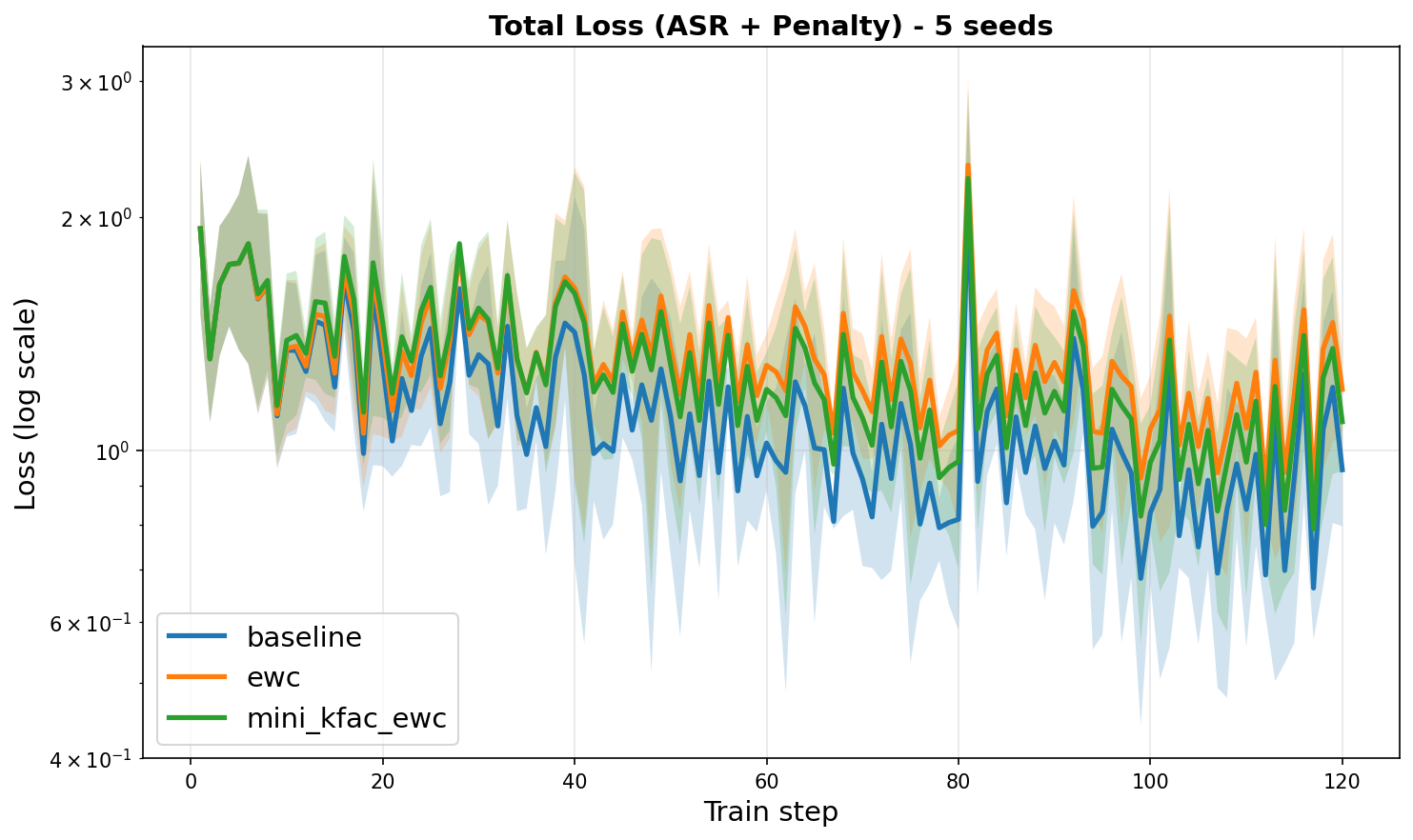}
\caption{Total training loss ($\mathcal{L}_{\text{total}} = \mathcal{L}_{ASR} + \mathcal{L}_{EWC}$) during Stage~B adaptation.}
\label{fig:total_loss}
\end{figure}

\begin{figure}[h]
\centering
\includegraphics[width=\linewidth]{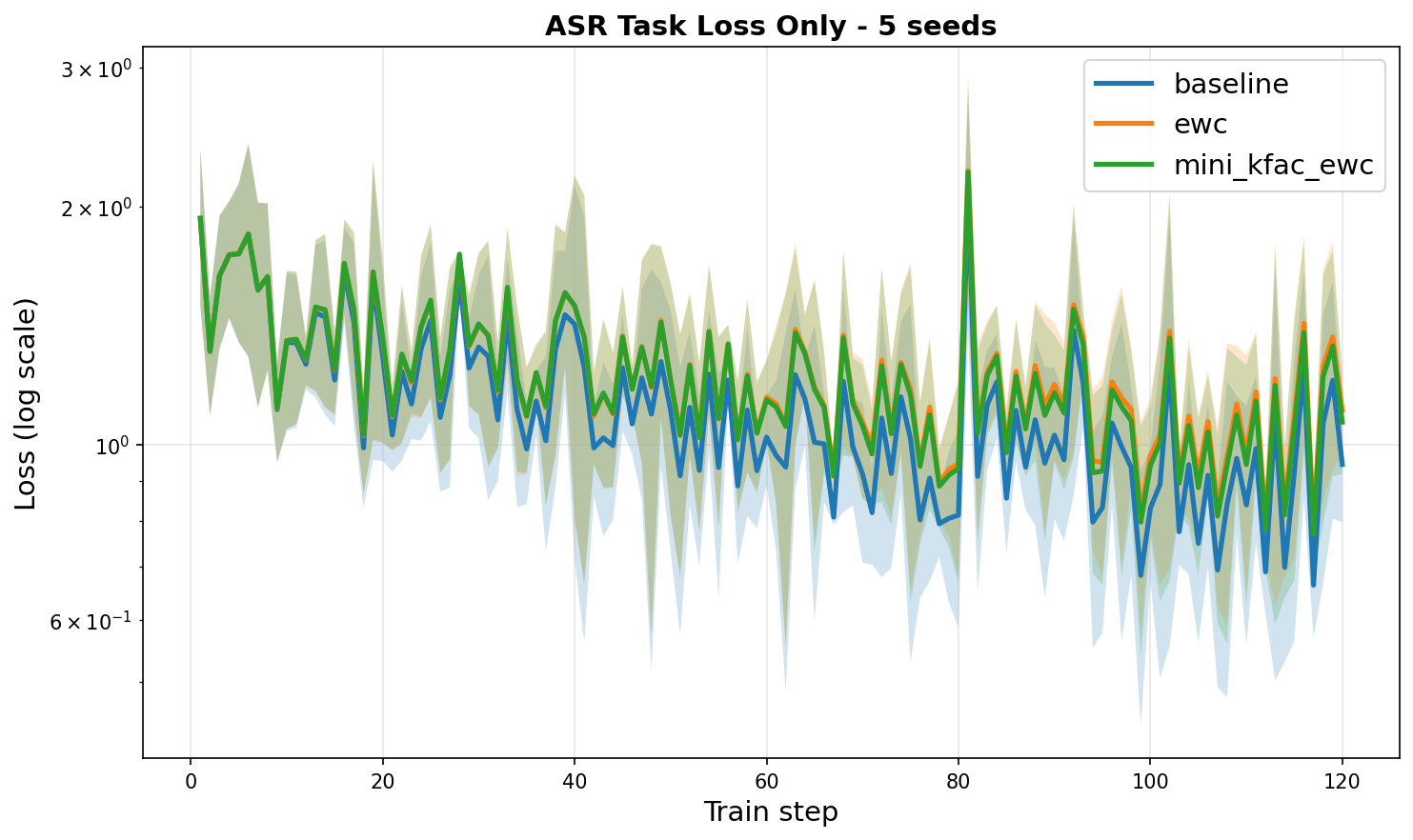}
\caption{ASR loss component ($\mathcal{L}_{ASR}$) during Stage~B adaptation.}
\label{fig:asr_loss}
\end{figure}

\begin{figure}[h]
\centering
\includegraphics[width=\linewidth]{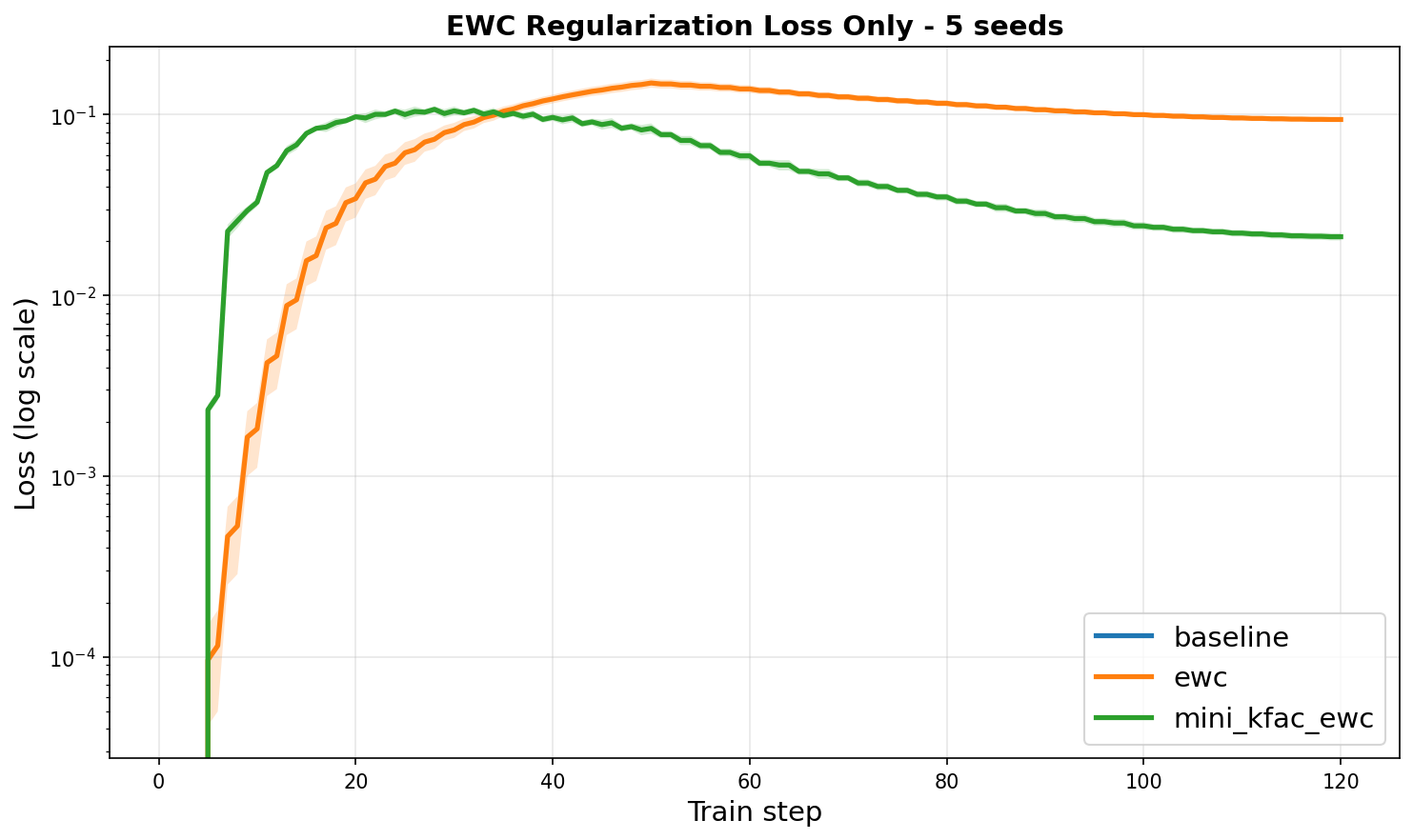}
\caption{EWC regularization loss component ($\mathcal{L}_{EWC}$) during Stage~B adaptation. The baseline shows zero EWC loss throughout, as no regularization is applied.}
\label{fig:ewc_loss}
\end{figure}

Figure~\ref{fig:penalty} shows the penalty dynamics during Stage~B fine-tuning. Both EWC methods rise initially due to two compounding effects: a 50-step linear ramp-up of the global scaling coefficient $\lambda$, from 0 to its target value, and growing parameter drift $(\theta - \theta^*)$ as the model adapts away from its Stage~A weights. These cause the effective penalty to increase even before the ramp completes. Mini-KFAC rises faster and peaks earlier as its Kronecker-factored structure detects structured parameter drift more precisely, while its penalty then decreases as the model finds weight-space directions satisfying both Task~B and the KFAC geometry. Diagonal EWC lacks this structure, accumulating penalty more uniformly and sustaining a higher, flatter trajectory throughout training.

The loss component breakdown (Figures~\ref{fig:total_loss}, \ref{fig:asr_loss}, and~\ref{fig:ewc_loss}): The same ramp-up dynamic explains the initial rise in total loss, $\mathcal{L}_{\text{total}}$, for regularized methods: as $\lambda$ scales up and parameter drift grows, the EWC term adds an increasing contribution to $\mathcal{L}_{\text{total}}$, constraining ASR improvement relative to the baseline. Mini-KFAC stabilizes at a lower, better-calibrated constraint level, allowing greater ASR progress while preserving retention, which explains its WER gains over diagonal EWC.

\section{Statistical Significance}
\label{app:stats}
\paragraph{Unit of analysis and test validity:}
The $t$-tests reported here rely on two assumptions: (1) the five WER values per method are approximately independent outcomes of repeated randomized training, and (2) they are approximately normally distributed. Both assumptions apply equally to the one-sample tests for negative forgetting (Table~\ref{tab:forgetting}) and the pairwise comparisons between methods (Table~\ref{tab:stats}), since both are computed from the same five per-seed WER values for each method. Both assumptions also differ from the concern raised in \citet{gillick1989some}, whose independence requirement addresses correlated word- or utterance-level errors within a single system's output on a single fixed test set, rather than independence across five separately-trained systems.

\textit{(1) Independence.} The five observations are approximately independent because each model is trained independently using a different random seed. Each seed determines the model initialization and dropout randomness, the batch order for both Stage A and Stage B training, which Stage A batches are used for Fisher estimation, and which rows are subsampled when estimating the mini-KFAC Kronecker factors (Appendix~\ref{app:hyperparams}). The evaluation partition is fixed and unshuffled, with Task A/B membership assigned deterministically from a hash of each file identifier rather than from the seed, so all five runs of a method are scored on the same utterances. Holding the evaluation set fixed removes evaluation-set variability and ensures that differences in WER primarily reflect differences induced by the randomized training procedure rather than differences in the data being evaluated. This supports treating the five WER measurements as approximately independent for both the one-sample and pairwise tests. It does not imply generalization beyond this particular validation partition, which is a separate question of external validity.

\textit{(2) Normality.} The reported $t$-tests assume that the five per-seed WER values are approximately normally distributed. Although this assumption cannot be verified empirically at $n=5$, its practical consequences differ between them. The one-sample tests for negative forgetting involve effect sizes that are large relative to the observed sampling variability (Table~\ref{tab:forgetting}, $p<0.05$ for all methods), so modest departures from normality are less likely to alter the qualitative conclusion than for the smaller pairwise differences. The pairwise comparisons between regularization strategies involve smaller observed differences, so the same unverifiable normality assumption is more consequential, and the tests correspondingly have limited power to detect true effects, consistent with the non-significant results in Table~\ref{tab:stats} despite directionally consistent trends across seeds. The findings therefore treat the pairwise comparisons as suggestive rather than conclusive, while placing more confidence in the one-sample negative-forgetting results. Bootstrap-based alternatives would help substantiate the pairwise comparisons further, since neither \citet{bisani2004bootstrap}'s confidence intervals nor \citet{liu2020statistical}'s blockwise bootstrap require a normality assumption, and both remain valid at small $n$.

\paragraph{Pairwise method comparisons:}
Table~\ref{tab:stats} reports Welch's two-sample $t$-tests comparing all method pairs across three metrics (Task~A WER, Task~B WER, Forgetting), with Cohen's $d$ as a complementary
effect size measure. No pairwise comparison reaches statistical significance at $n=5$.
The mini-KFAC vs.\ diagonal EWC differences are small in both absolute and effect-size
terms ($d \leq 0.23$), confirming that stronger conclusions about relative regularizer
quality require more seeds or longer task sequences. The baseline advantage over regularized
methods, while not significant, exhibits large effect sizes on Task~B WER ($d \approx 0.83$--$0.89$),
suggesting a real pattern that larger experiments could confirm.

\begin{table}[h]
\centering
\caption{Welch's two-sample $t$-tests ($n=5$ seeds per method, two-tailed).
$d$ = Cohen's $d$ (pooled std). No comparison is significant at $\alpha=0.05$.}
\label{tab:stats}
\small
\resizebox{\columnwidth}{!}{
\begin{tabular}{llccccc}
\toprule
\textbf{Comparison} & \textbf{Metric} & $\boldsymbol{\Delta}$ \textbf{(\%)} & $\boldsymbol{t}$ & \textbf{df} & $\boldsymbol{p}$ & $\boldsymbol{d}$ \\
\midrule
mini-KFAC vs EWC & Task A WER   & $-0.36$ & $-0.27$ & $8.0$ & $0.796$ & $-0.17$ \\
                 & Task B WER   & $-0.97$ & $-0.36$ & $7.3$ & $0.731$ & $-0.23$ \\
                 & Forgetting   & $-0.36$ & $-0.26$ & $7.7$ & $0.802$ & $-0.16$ \\
\midrule
Baseline vs EWC  & Task A WER   & $-1.20$ & $-1.02$ & $7.6$ & $0.341$ & $-0.64$ \\
                 & Task B WER   & $-3.29$ & $-1.41$ & $5.0$ & $0.217$ & $-0.89$ \\
                 & Forgetting   & $-1.21$ & $-0.86$ & $7.6$ & $0.416$ & $-0.54$ \\
\midrule
Baseline vs mini-KFAC & Task A WER & $-0.84$ & $-0.69$ & $7.4$ & $0.510$ & $-0.44$ \\
                      & Task B WER & $-2.32$ & $-1.31$ & $5.8$ & $0.241$ & $-0.83$ \\
                      & Forgetting & $-0.85$ & $-0.55$ & $8.0$ & $0.597$ & $-0.35$ \\
\bottomrule
\end{tabular}
}
\end{table}

\paragraph{Negative forgetting (task complementarity):}
To test whether forgetting is significantly negative which means whether Task~A performance genuinely improves during Stage~B, one-sample $t$-tests are applied against the null hypothesis of zero forgetting ($H_0$: $\text{Forgetting} \geq 0$), using one-tailed tests:

\[
t = \frac{\bar{x}_{\text{forgetting}}}{s / \sqrt{n}}, \quad df = n - 1 = 4
\]

\begin{table}[h]
\centering
\caption{One-sample, one-tailed $t$-tests against $H_0$: Forgetting $\geq 0$, $n=5$ seeds per method, $df=4$.}
\label{tab:forgetting}
\resizebox{\columnwidth}{!}{
\begin{tabular}{lccc}
\toprule
\textbf{Method} & $\boldsymbol{t}$ & $\boldsymbol{p}$ \textbf{(one-tailed)} & \textbf{Sig.} \\
\midrule
Baseline       & $-4.41$ & $0.006$ & $p < 0.01$ \\
EWC (Diagonal) & $-4.20$ & $0.007$ & $p < 0.01$ \\
Mini-KFAC EWC  & $-3.73$ & $0.010$ & $p < 0.05$ \\
\bottomrule
\end{tabular}
}
\end{table}

All three methods show statistically significant negative forgetting ($p < 0.05$, one-tailed), confirming that task complementarity in financial speech is a robust empirical phenomenon and not an artifact of small sample variance.

\section{Gradient Flow Analysis}
\label{app:gradients}

To monitor the interaction between the ASR objective and EWC regularization throughout Stage~B training, gradient magnitude ratios are tracked:
\[
\text{grad\_ratio} = \frac{\|\nabla_\theta \mathcal{L}_{EWC}\|}{\|\nabla_\theta \mathcal{L}_{ASR}\| + \epsilon}
\]
Regularized methods maintain stable non-zero ratios after the 50-step ramp-up, confirming that the EWC term contributes consistently to $\mathcal{L}_{\text{total}} = \mathcal{L}_{ASR} + \mathcal{L}_{EWC}$ rather than vanishing or dominating. The baseline has zero EWC gradient throughout, allowing unconstrained optimization toward the lowest ASR loss. Moreover, smooth ratio curves indicate stable optimization dynamics for all methods.

\section{When These Findings May Not Apply}
\label{app:when_reg_works}

These results are obtained in a setting where Task~A and Task~B are closely related financial-speech tasks and exhibit negative forgetting. Different conclusions may arise when tasks are less aligned. In particular, continual-learning regularization may be beneficial under three conditions: (1) substantial acoustic or linguistic mismatch between tasks, (2) strong catastrophic forgetting during sequential adaptation, or (3) limited Task~B training data, where regularization can help preserve prior knowledge and reduce overfitting. These scenarios remain important directions for future evaluation.

\section{Future Directions}

\begin{enumerate}
    \item Task similarity assessment: Develop quantitative measures of task relatedness in financial speech (representation similarity, Fisher spectrum alignment, loss overlap) to predict when regularization is beneficial.
    \item Adaptive regularization: Design dynamic $\lambda$ weighting that decreases when Task~A performance improves during Stage~B adaptation, enabling more flexible transfer in complementary domains.
    \item Selective layer-wise constraints: Investigate regularization of only layers with high task specificity through gradient attribution or attention sensitivity, rather than global Fisher penalties.
    \item Longer financial sequences: Extend evaluation to multi-stage streams (for example, earnings calls across fiscal quarters, investor presentations, and regulatory transcripts) to determine whether complementarity persists.
    \item Broader baselines: Compare Fisher-based methods against replay, adapter-based fine-tuning, and progressive architectures to situate mini-KFAC EWC within the broader continual learning landscape for financial ASR.
\end{enumerate}

\end{document}